\documentclass[reprint,10pt,superscriptaddress,twocolumn,aps,prx]{revtex4-2}
\usepackage{amsmath}
\usepackage{amssymb}
\usepackage{graphicx}
\usepackage{physics}
\usepackage{bbm}
\usepackage{color}
\usepackage{dcolumn}% Align table columns on decimal point
\usepackage[colorlinks=true,citecolor=blue,linktocpage=true,linkcolor=blue,filecolor=blue,urlcolor=magenta]{hyperref}
\usepackage{physics}
\usepackage{bbm}
\usepackage{color}
\usepackage{dcolumn}% Align table columns on decimal point

\makeatletter
\def\maketitle{
\@author@finish
\title@column\titleblock@produce
\suppressfloats[t]}
\makeatother

\begin{document}

\preprint{APS/123-QED}

\title{Image Current Detection of Electrons in a Room-Temperature Paul Trap}% Force line breaks with \\

\author{Kento Taniguchi}
 \email{kento-taniguchi531@berkeley.edu}
 \altaffiliation[Present address: ]{Department of Physics, University of California, Berkeley, Berkeley, CA 94720, USA}
 \affiliation{Komaba Institute for Science (KIS), The University of Tokyo, Meguro-ku, Tokyo, 153-8902, Japan}%

\author{Atsushi Noguchi}
\email{u-atsushi@g.ecc.u-tokyo.ac.jp}
\affiliation{Komaba Institute for Science (KIS), The University of Tokyo, Meguro-ku, Tokyo, 153-8902, Japan}
\affiliation{RIKEN Center for Quantum Computing (RQC), RIKEN, Wako-shi, Saitama 351-0198, Japan}
\affiliation{Inamori Research Institute for Science (InaRIS), Kyoto-shi, Kyoto, 600-8411, Japan}%

\date{\today}% It is always \today, today,

\begin{abstract}
We report the image current detection of electrons in a room-temperature Paul trap at microwave frequencies. By selectively leveraging distinct cavity modes for trapping and detection, our approach efficiently extracts electron signals otherwise buried in the microwave drive used for pseudo-potential formation. When the trapped electrons resonate with the cavity mode, we observe a mode excitation and its exponential decay attributed to resistive cooling. Detuning electrons from the cavity resonance halts this decay, and sweeping electrons' secular frequency reveals their oscillatory spectrum. Implementing this experiment at cryogenic temperatures could enable the image current detection and ground-state cooling of a single electron in Paul traps.  
\end{abstract}

\maketitle

%\tableofcontents

\section{Introduction} Electrons trapped in vacuum provide an exceptionally low-noise quantum system, enabling ultra-precise experiments. In particular, Penning traps have realized the most precise measurements of the electron $g$-factor, providing a stringent test of the Standard Model \cite{Fan2023}. Beyond fundamental physics, these systems have demonstrated the Purcell effect \cite{Gabrielse1985, Hanneke2011}, resolved side-band cooling \cite{wineland1975, Brown1986}, and quantum non-demolition measurements of spin states and cyclotron motion \cite{van1976, Peil1999}. These achievements have laid a solid foundation for quantum physics, which developed into various quantum technologies \cite{Blais2021, Monroe1995, Hamann1998, Braginsky1996}.

Trapped electrons are also advantageous platforms for quantum information processing. Their light mass enables motional frequencies in the gigahertz range, allowing for faster quantum operations compared to trapped ions \cite{haffner2008}. The spin state of a trapped electron, an isolated pure two-level system, can be a robust qubit resistant to quantum information leakage out of the computation subspace. Moreover, both motional and spin states can be controlled using well-established microwave technologies, which readily support microfabrication-based architecture. Building on these benefits, various quantum information processing schemes have been proposed \cite{Mancini1999, Ciaramicoli2001, Ciaramicoli2003, Ciaramicoli2004, Ciaramicoli2007, Ciaramicoli2008, Zurita2008, Marzoli2009, Lamata2010, Ciaramicoli2010, bushev2011, Rosales2018, Cridland2020}, expanding the applications of trapped electrons from fundamental physics to quantum computation.

Various trapping mechanisms have been explored to implement electron-based qubits in Penning traps. Planar Penning traps have gained attention due to their potential for two-dimensional scalability \cite{stahl2005, galve2006, galve2007}, which is required for quantum error correction algorithms such as surface codes \cite{Fowler2012}. However, their inherent lack of symmetry introduces significant anharmonicity in the trap potential, making image current detection of a single electron infeasible \cite{bushev2008}. While theoretical methods have been proposed to mitigate this issue \cite{Goldman2010}, no successful single electron measurements have been reported.

Unlike Penning traps, where an electron exhibits large-radius magnetron motion, Paul traps confine an electron closer to the trap center where anharmonicity is minimal. This localization enhances the feasibility of both single electron detection and scalable two-dimensional architectures, shifting the attention toward trapping electrons in Paul traps \cite{Daniilidis2013, Kotler2017, Peng2017, Qian2022, huang2025}. This was recently demonstrated 
 at room temperature \cite{Matthiesen2021}, with detection achieved by ejecting trapped electrons onto a micro-channel plate. 

Essentially, quantum computation demands repeated operations, requiring the continuous confinement of electrons and their image current detection \cite{Brown1986}. Equally crucial is ground-state cooling of the electron's motion, a prerequisite for precise quantum manipulations \cite{Osada2022}. However, both approaches present challenges unique to Paul traps: the former requires stringent filtering of the trap's microwave field to extract the electron signal and protect detection circuits \cite{Qian2022}, while the latter demands resistive cooling at cryogenic temperatures, necessitating Paul traps inside a cryostat where cooling capacity is limited \cite{Matthiesen2021}. Overcoming these obstacles requires an integrated experimental framework combining advanced microwave filtering with power-efficient trap designs. 

\begin{figure*}[ht]
\includegraphics[width=16.8cm]{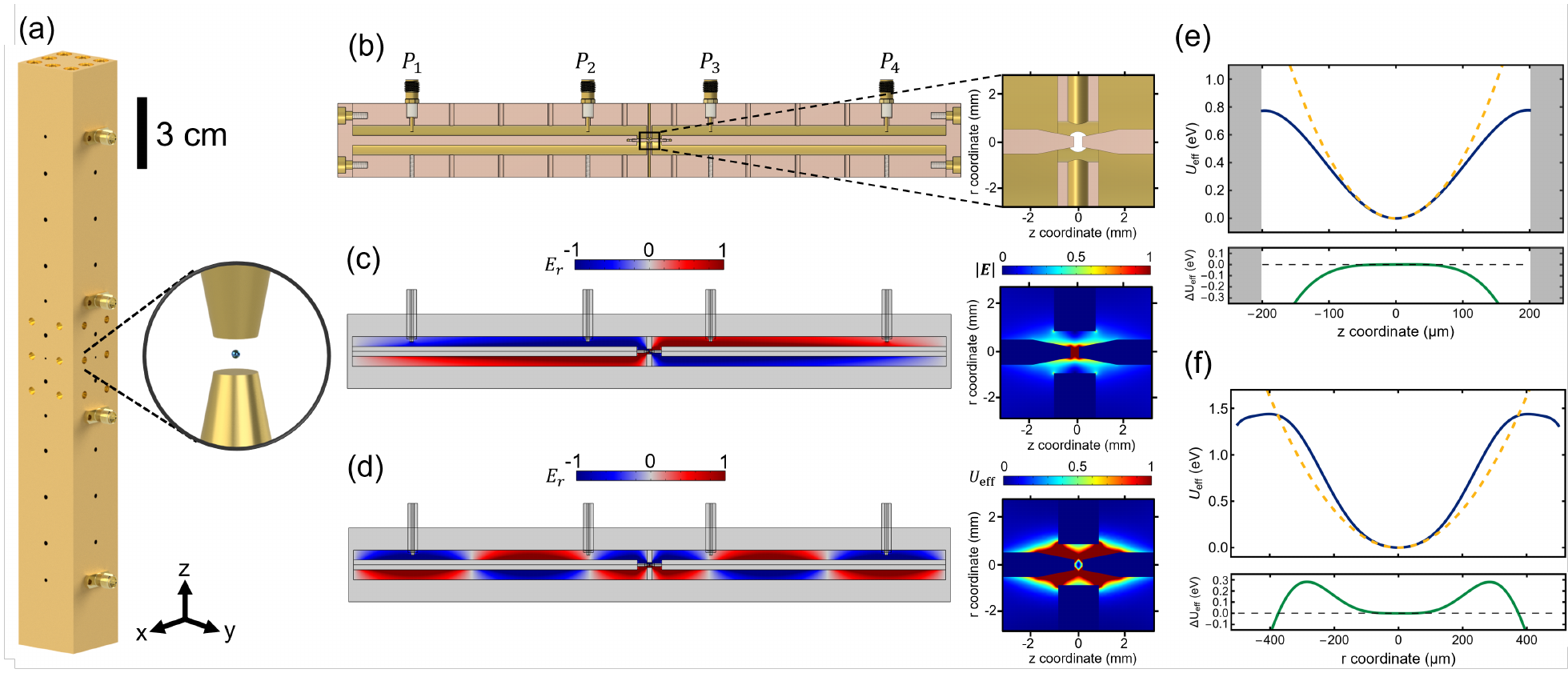}
\caption{\label{Fig1} Schematic illustration and characterization of the three-dimensional coupled coaxial cavities. (a) Design of the three-dimensional coupled coaxial cavities. (b) Cross-sectional view of the coupled coaxial cavities. The inset provides a magnified view of the pin geometry, where electrons are confined. (c) Radial ($r$-axis) electric field distribution of the out-of-phase $\lambda/4$ mode (readout mode) in cylindrical coordinates, used for image current detection of trapped electrons. The inset highlights the norm of electric fields between the pins, forming a localized electric dipole field. (d) Radial electric field distribution of the in-phase $5\lambda/4$ mode (trapping mode), which generates the pseudo-potential. The inset illustrates the corresponding potential structure responsible for electron confinement. (e) Top: Pseudo-potential $U_\mathrm{eff}$ along the $z$-axis, centered at potential minimum (blue curve), compared to an ideal harmonic potential (dashed orange curve). Bottom: Deviation $\Delta U_\mathrm{eff}$ from the ideal harmonic potential, quantifying the anahronicity. The gray regions indicate the positions of the pin tips. (f) Pseudo-potential distribution and its deviations along the radial axis which reveal the influence of the cavity geometry on the potential shape.}
\end{figure*}

To address these challenges, we previously proposed three-dimensional coupled coaxial cavities \cite{Osada2022}. This device can efficiently extract electron signals from the trap's microwave field by leveraging distinct resonance modes for pseudo-potential formation and image current detection. Moreover, this cavity concentrates electric fields at a single point and has a very high $Q$ factor when integrated with superconducting materials. Consequently, their synergy facilitates power-efficient electron Paul traps ($\sim \mathrm{mW}$) compatible with cryogenic operations. In this work, we successfully realize electron Paul traps using the coupled coaxial cavities. We achieve image current detection of electrons and observe exponential signal decay attributed to resistive cooling at room temperature. These results pave the way towards realizing image current detection and ground-state cooling of a single electron qubit at cryogenic temperature.

\section{Method}
\subsection{Three-dimensional Coupled Coaxial Cavities}
We begin by introducing the three-dimensional coupled coaxial cavities (see Fig.~\ref{Fig1}). This device consists of two open-end $\lambda /4$ coaxial cavities capacitively coupled through pins placed on central pillars [Fig.~\ref{Fig1}(b)]. This configuration generates in-phase and out-of-phase modes with wavelengths of around $(2n+1)\lambda/4 \ \forall n \in \mathbb{N}^+$. The out-of-phase mode generates an electric dipole field between the pins [Fig.~\ref{Fig1}(c)], facilitating the dipole interaction between an intracavity photon and the secular motion of trapped electrons. Conversely, the higher-order in-phase mode produces a quadratic electric potential between the pins, which is suitable for the electron Paul trap [Fig.~\ref{Fig1}(d)]. Consequently, electrons can be trapped between the pins using high-order in-phase mode and electrically detected with low-order out-of-phase modes.

We fabricated three-dimensional coupled coaxial cavities made of gold-plated copper, as shown in Fig.~\ref{Fig1}(b). This device has many holes to ensure the vacuum quality inside, which does not degrade the quality factor as long as the boundary condition is maintained \cite{Chakram2021}. SMA panel mount connectors are inserted into the cavities as coupling ports, strategically positioned to induce trap microwave and extract the signal of trapped electrons, as illustrated in Fig.~\ref{Fig1}(b-d). Ports 2 and 3 ($P_2$ and $P_3$), placed at the node of the in-phase $5 \lambda/4$ mode (trapping mode), are utilized for detecting trapped electrons. These ports are coupled with the out-of-phase $\lambda/4$ mode (readout mode) but only weakly interact with the trapping mode, reducing the leakage of the trap microwave into the detection apparatus, such as amplifiers and spectrum analyzers. Conversely, ports 1 and 4 ($P_1$ and $P_4$) are positioned at the antinode of the trapping mode [Fig.~\ref{Fig1}(d)]. This maximizes the coupling between the trapping mode and the ports, allowing its efficient excitation and the pseudo-potential formation. 

The transmission and reflection spectrum of the fabricated cavity revealed an internal quality factor of $Q_\mathrm{in} \approx 1,300$ and an external quality factor of $Q_\mathrm{ex} \approx 20,000$ with resonance frequency of $\omega_1^\mathrm{out}/2 \pi \approx 619 \ \mathrm{MHz}$ for the readout mode, and $Q_\mathrm{in} \approx 2,300$ and $Q_\mathrm{ex} \approx 3,300$ at the resonance frequency of $\omega_5^\mathrm{in}/2\pi \approx 3.105 \ \mathrm{GHz}$ for the trapping mode. This configuration enables the electron Paul trap and image current detection with $q \approx 0.56$ stability parameter. Despite efforts to maximize the external coupling by positioning the central pin of the SMA connector as close to the pillar as possible, the readout mode remained under coupling due to the position constraints in locating at the node of the trapping mode. Employing the superconducting cavity at cryogenic temperature would achieve a higher $Q_{\mathrm{in}}$, resolving this issue.

The image current detection of trapped electrons involves coupling their oscillations to the intracavity photons of three-dimensional coupled coaxial cavities via electrical dipole interaction. This interaction is described by the Hamiltonian:
\begin{equation}
    H = \hbar g \left(\hat{a}_e^\dag \hat{a}_p + \hat{a}_e \hat{a}_p^\dag \right), 
\end{equation}
where $\hat{a}_e$ and $\hat{a}_p$ are annihilation operators of an electron motional phonon along $z$-axis and an intracavity photon, and $g/2\pi$ is their coupling coefficient. The radial motional phonon can also be controlled by coupling them to $\hat{a}_e$ \cite{Gorman2014}. Using Finite-Element simulation with COMSOL, we estimate the coupling coefficient to be $g/2\pi \approx 54 \ \mathrm{kHz}$ for an electron with secular frequency $\omega_z/2\pi = \omega_1^\mathrm{out}/2\pi = 619 \ \mathrm{MHz}$.

The pseudo-potential for the electron Paul trap is generated by oscillating quadrupole electric potential and can be formulated as $U_\mathrm{eff} = (q_e^2 E^2 /4m_e \Omega^2)$. Here, $E$ and $\Omega$ represent the amplitude and angular frequency of the oscillating inhomogeneous electric field, while $q_e$ and $m_e$ are the charge and mass of the electron. Figure~\ref{Fig1}(e) and \ref{Fig1}(f) show simulated pseudo-potentials along r- and z-axes when the pin voltages are $92\, \mathrm{V}$, which sets the electron oscillation to resonate with readout mode ($\omega_z = \omega_1^\mathrm{out}$). This pin voltage can be achieved with a microwave input power of $1.5\, \mathrm{W}$ under the current coupling condition and could be reduced to $800\, \mathrm{mW}$ if the critical coupling is achieved. The potential depth is approximately $0.8\, \mathrm{eV}$, limited by the confinement along the $z$-direction. The anharmonicity of the trap potential is illustrated as $\Delta_\mathrm{eff}$, the deviations from the ideal harmonic potentials (shown as orange dashed lines) in the bottom panels highlighted as green curves.

\begin{figure*}[ht]
\includegraphics[width=16.8cm]{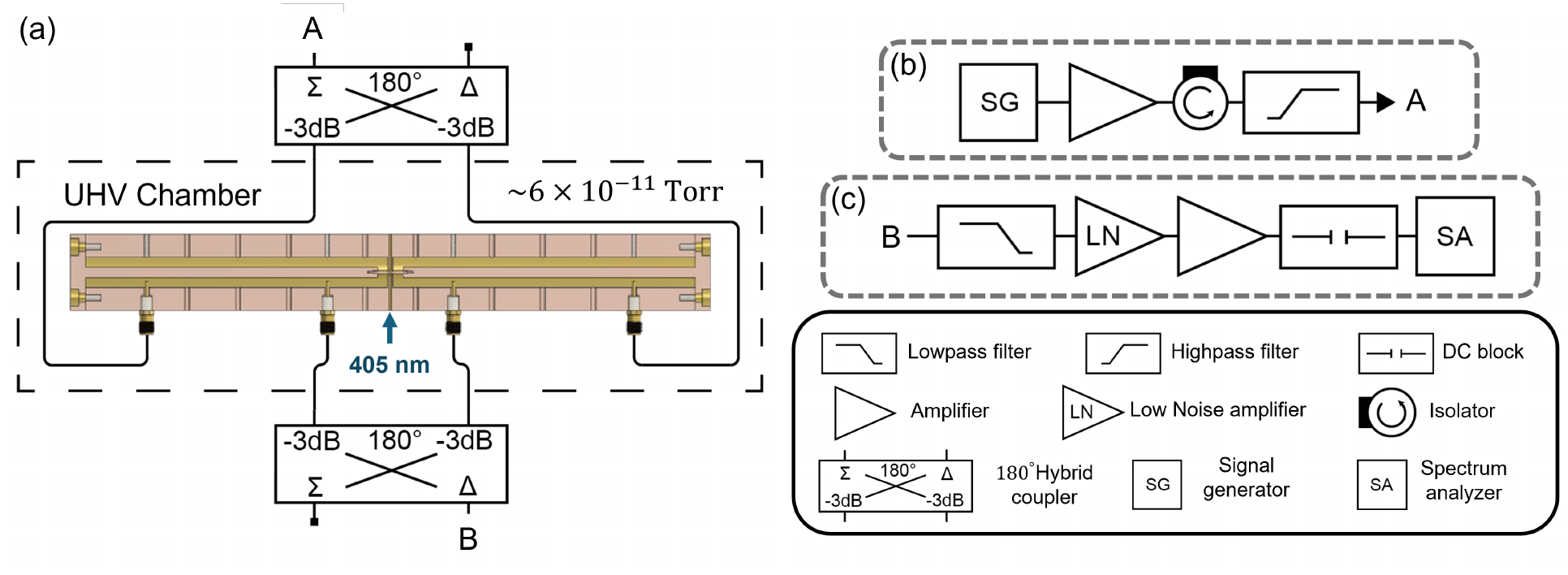}
\caption{\label{Fig2} Experimental setup. (a) Connections between the three-dimensional coupled coaxial cavities inside the UHV chamber to ports A and B. (b) Trapping microwave generation systems for pseudo-potential formation. (c) Measurement system for the image current detection of trapped electrons.}
\end{figure*}

We note that anharmonicity in the Pseudo-potential causes amplitude-dependent frequency shifts in the secular motion of trapped electrons. This effect broadens the motion spectrum and degrades the signal-to-noise ratio (SNR) in image current detection \cite{bushev2008}. In our coupled coaxial cavities, anharmonicity primarily depends on the separation and the radius of the pin electrodes. For our experiment, we used a pin separation of $400\,\mathrm{\mu m}$ and a radius of $250 \, \mathrm{\mu m}$. Considering the symmetry of the system, the pseudo-potential along the $z$-axis can be modeled using $U_\mathrm{eff}(z) = \frac{1}{2}m_e \omega_z^2 (z^2 + C_4 z^4 + C_6 z^6)$ ignoring odd-order terms. From the polynomial fit to the potential shown in Fig.~\ref{Fig1}(e), we obtained $C_4 = -1.5\cross 10^{-5} \, \mathrm{\mu m}^{-2}$ and $C_6 \approx 0 \, \mathrm{\mu m}^{-4}$. Therefore, the amplitude-dependent secular frequency shifts are given by \cite{Goldman2010}:
\begin{equation}
    \omega_z (z) \approx \omega_z \left[1 + \frac{3 C_4}{4} z^2 - \frac{21 (C_4)^2}{64} z^4 + \cdots \right].
    \label{amplitude_dependent_shifts}
\end{equation}
Here, $\omega_z (z)$ is the secular frequency at amplitude $z$, and $\omega_z$ is the secular frequency when there is no anharmonicity. The secular frequency shifts $\Delta \omega_z = |\omega_z(z) - \omega_z|$ approximately corresponds to the bandwidth of the electron motion spectrum when broadening is dominated by anharmonicity. For $\omega_z = 2 \pi \cross 619 \ \mathrm{MHz}$ and the thermal spread of trapped electron at $T = 300 \ \mathrm{K}$ with $z = \sigma_\mathrm{std} \approx 17.3 \ \mathrm{\mu m}$, the frequency shift is $\Delta \omega_z \approx 2 \pi \cross 2 \ \mathrm{MHz}$. This broadening significantly exceeds the cooling rate of the electron's oscillation $\gamma_z = 4 g^2/\kappa \approx 2 \pi  \cross 24 \ \mathrm{kHz}$, where $\kappa \approx 2 \pi \cross 476 \ \mathrm{kHz}$ denotes the cavity photon loss rate. As a result, SNR for single-electron detection is reduced by more than an order of magnitude, preventing reliable single-electron measurement. Therefore, this experiment focuses on demonstrating the proof of concept for the image current detection of multiple electrons in Paul traps with coupled coaxial cavities.

\subsection{Electron Source}
To efficiently load electrons into traps, the kinetic energy of the loaded electrons must be sufficiently lower than the trap depth, highlighting the importance of preparing a low-energy electron source. In the Penning trap system, thermionic or field emission points are frequently used for this purpose \cite{richardson1929, Robertson2000}. In these methods, high-energy primary electrons are generated first and collide with the device wall, producing background gases, and those background gases collide with other high-energy primary electrons, producing lower-energy secondary electrons that are supposed to be trapped. 

This study used the photoelectron effect to generate high-energy primary electrons. We directed a continuous multimode laser with a central wavelength of $405 \, \mathrm{nm}$ and a power of $300 \, \mathrm{mW}$ onto the pins inside the coupled coaxial cavities through a hole [see Fig.~\ref{Fig2}(a)]. Although the energy of the laser at $405 \, \mathrm{nm}$ (approximately $3.06 \, \mathrm{eV}$) is theoretically insufficient to produce photoelectrons given the work function of gold plated on the device wall ($5.30 \, \mathrm{eV}$) \cite{SACHTLER1966}, biasing the pins with an oscillating electric field enables Schottky-enabled photoemission \cite{Yusof2004}. In addition, the multiple-photon process also facilitates photoemission \cite{Ueda2007}.

\subsection{Experimental Setup}
The configuration of the electron Paul trap experiment is described in Fig.~\ref{Fig2}(a). The coupled coaxial cavities are placed inside the Ultra-High-Vacuum (UHV) chamber with pressure below $6\cross 10^{-11}\, \mathrm{Torr}$. The photoemission laser for electron loading is introduced into the chamber through an optical window. 

The pseudo-potential of the electron Paul trap is formed by the continuous microwave excitation of in-phase $5\lambda/4$ mode (trapping mode) of coupled coaxial cavities. This trap microwave (trap MW), operating at a frequency of around $3.105 \, \mathrm{GHz}$, is first amplified using a high-power amplifier and then passed through an isolator and a high-pass filter with a cut-off frequency of $1.9 \, \mathrm{GHz}$ to protect amplifier from reflection damages and to prevent the leakage of the amplifier noise into the out-of-phase $\lambda/4$ mode (readout mode) [Fig.~\ref{Fig2}(b)]. For selective excitation of the trapping mode, the trap MW is split using a Double-Arrow $180^\circ$ Hybrid Coupler and inputted into the three-dimensional coupled coaxial cavities via coupling ports $P_1$ and $P_4$ with the same phase. To eliminate any relative phase shifts between them, the lengths of wires from the hybrid coupler to the cavities were tuned the same.

The signal of trapped electrons was observed by measuring the excitation of the readout mode via coupling ports $P_2$ and $P_3$ using a spectrum analyzer. Here, filtering the trap MW field from the electron signals is crucial to protect the amplifier and the spectrum analyzer. Therefore, we implemented a three-stage filtering in the electronic circuit. Firstly, coupling ports $P_2$ and $P_3$ are positioned at the node of the trapping mode, resulting in a $30 \, \mathrm{dB}$ reduction of trap MW leakage into the detection circuit [Fig.~\ref{Fig1}(d)]. Secondly, coupling ports $P_2$ and $P_3$ are connected to the Double Arrow $180^\circ$ Hybrid Coupler to induce destructive interference of the trap MW [Fig.~\ref{Fig2}(a)]. While the trapping mode excites these ports in phase, the readout mode excites them with a phase difference of $\pi$. Since the hybrid coupler processes the input signals by outputting their sum or difference, an appropriate choice of output ports can suppress the signal from the trapping mode by $16 \, \mathrm{dB}$ while preserving the electron signals. Finally, the signal is transmitted through two low-pass filters with a cut-off frequency of $1.65 \, \mathrm{GHz}$, eliminating the trap MW by $80 \, \mathrm{dB}$ [Fig.~\ref{Fig2}(c)]. These result in a total filtering of approximately $126 \, \mathrm{dB}$ for trap MW. After the filtering, the readout mode excitation signals are amplified by $62\,\mathrm{dB}$ with a low-noise amplifier followed by another amplifier before the signal analyzer [Fig.~\ref{Fig2}(c)].

\section{Results}
\subsection{Image Current Detection}
\begin{figure}[ht]
\includegraphics[width=8.4cm]{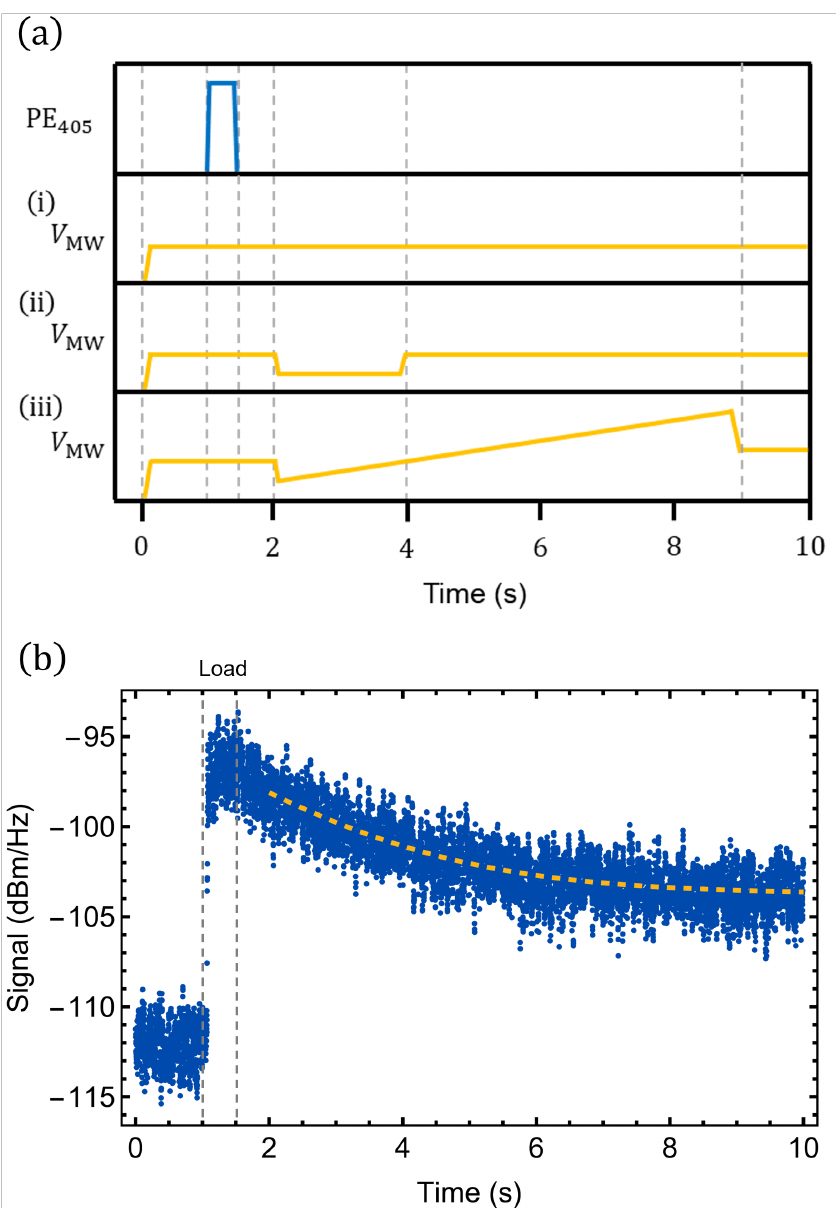}
\caption{\label{Fig3} Measurement sequences and observed signal. (a) Electrons are loaded after the trap MW is activated, with a $405 \, \mathrm{nm}$ laser illuminating the pins. This is followed by three measurement sequences:(i) The trap MW amplitude remains constant. (ii) The trap MW amplitude is temporarily varied, detuning the trapped electrons' COM mode from the cavity and then returning it. (iii) The trap MW amplitude is modulated to sweep electrons' COM frequency across the cavity resonance spectrum. (b) The readout mode excitation observed in sequence (i) is shown as a blue line. An exponential fit (dashed orange curve) yields a decay time of $1.74 \, \mathrm{s}$.}
\end{figure}

The experimental sequence is illustrated in Fig.~\ref{Fig3}(a). It begins with the activation of the trap MW at $0\, \mathrm{s}$, followed by the laser illumination for electron loading at $1\mathrm{s}$ for $0.5$ seconds, along with specified $V_{\mathrm{MW}}$ modulations. Throughout the sequence, the excitation of the readout mode is continuously monitored using a spectrum analyzer with zero-span mode. 

When the $V_{\mathrm{MW}}$ was appropriately tuned to resonate electrons' center-of-mass (COM) motion with the readout mode, an excitation of the readout mode was observed. Figure~\ref{Fig3}(b) presents the results for sequence (i), where $V_{\mathrm{MW}}$ remains constant to keep resonating electrons with the readout mode. The excitation signal appeared and gradually decayed with a time constant of $t_{\mathrm{cool}} = 1.74 \ \mathrm{s}$ before reaching a steady state. The observed excitation lasted over ten seconds even after electron loading finished, confirming the successful Paul trap of electrons and their image current detection.

Notably, both the initial excitation magnitude and the lifetime of trapped electrons exhibit variations between experimental cycles, generally decreasing with repeated trials. This degradation could stem from either a decline in vacuum quality within the coaxial cavity, possibly due to limited conductance and outgassing induced by the trap MW, or an increment in stray electric fields caused by stray charges accumulated on electrode contamination.

\subsection{Secular Frequency Modulation}
The observed signal decay can arise from either resistive cooling or the loss of trapped electrons. In resistive cooling, the signal should dissipate through the resistive losses of the cavity and detection circuit but can be halted by detuning the electrons from resonance. In contrast, electron loss leads to signal dissipation as electrons escape the trap, independent of resonance conditions. To distinguish these mechanisms, we conducted an experiment in which trapped electrons were successively tuned on- and off-resonance with the readout mode, as shown in sequence (ii) of Fig.~\ref{Fig3}(a).

\begin{figure}[ht]
\includegraphics[width=8.4cm]{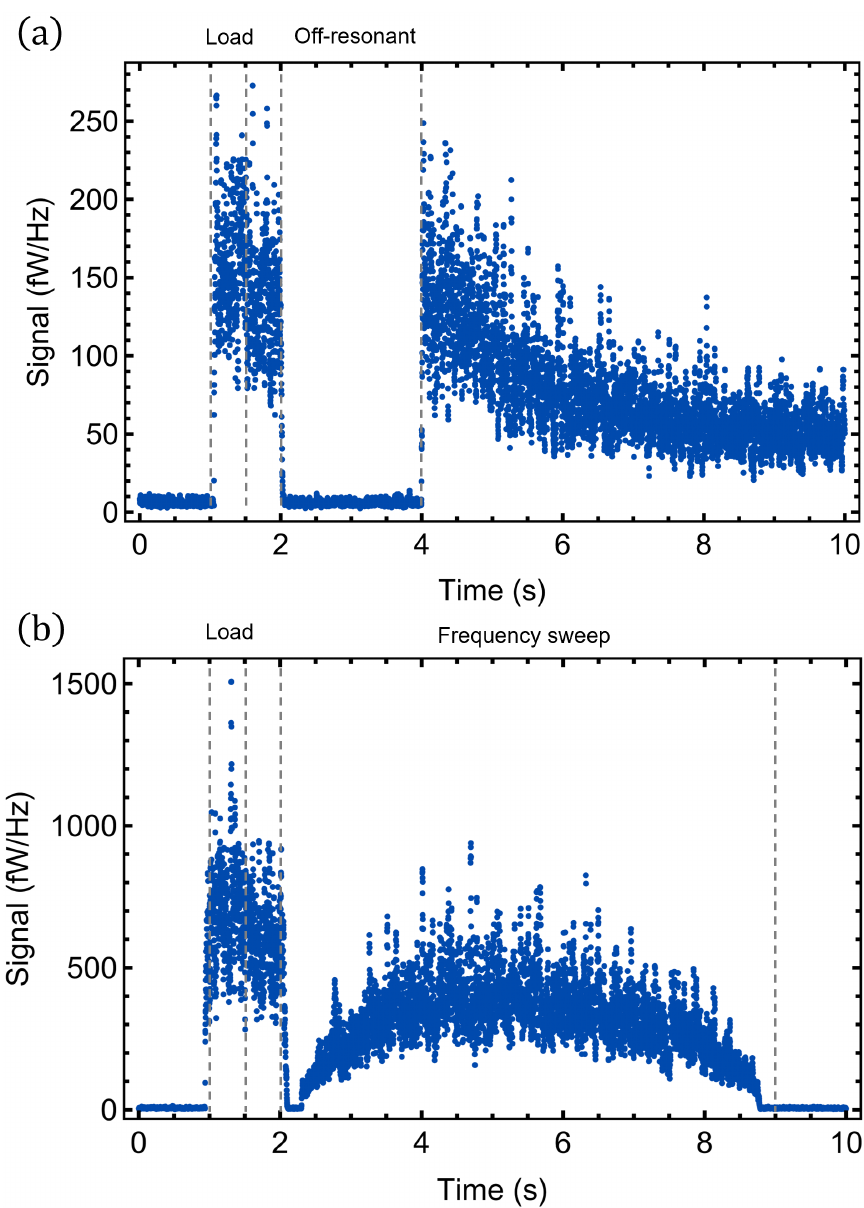}
\caption{\label{Fig4} Secular frequency modulation and oscillatory spectrum. (a) The readout mode excitation is due to trap MW amplitude modulation in sequence (ii). The signal decay stopped when electrons moved off-resonance with the readout mode, suggesting resistive cooling. (c) That of sequence (iii), where electrons' COM mode secular frequency is swept across the readout mode resonance spectrum. The resonant excitation of the cavity mode was observed.}
\end{figure}

In the experiment with sequence (ii), we modulated the $V_\mathrm{MW}$ amplitude to shift the COM mode frequency of the trapped electrons by approximately $62 \, \mathrm{MHz}$ at around $2\,\mathrm{s}$ for a duration of $2\,\mathrm{s}$, temporarily detuning them from the readout mode before restoring resonance. Figure~\ref{Fig4}(a) presents the excitation of the readout mode during this process, showing the disappearance and subsequent revival of the signal in response to changes in the $V_\mathrm{MW}$ amplitude. The fully restored signal after detuning indicates that the decay ceases when the electrons are off-resonance, confirming that resistive cooling is the primary cause of the observed signal decay. 

However, the measured cooling rate, $\gamma_\mathrm{ex}/2\pi = 0.57 \, \mathrm{Hz}$, is significantly slower than the expected rate of $\gamma_z/2 \pi = 24 \, \mathrm{kHz}$. Three factors could contribute to this discrepancy. First, anharmonicity in the pseudo-potential may induce fluctuations in the COM frequency, reducing the efficiency of resistive cooling. Second, weak coupling between the COM mode and other radial motions, caused by trap nonlinearity and Coulomb interactions, may limit the total cooling rate, as observed in previous Penning trap experiments \cite{Kiffer2024}. Third, parametric excitation of the COM mode by micromotion could degrade the cooling efficiency \cite{Van2022}. In Paul traps, the strong microwave drive used to generate the pseudo-potential continuously excites micromotion, and due to the system's nonlinearity, this could lead to persistent excitation of the COM mode, counteracting cooling. 

Further systematic studies are required to fully understand these effects. However, the significant fluctuations in the decay rate and secular frequency observed in our current setup prevent it. The pseudo-potential anharmonicity amplified by the large thermal spread of electrons at room temperature ($300 \, \mathrm{K}$) seems to play a major role in this instability. Therefore, future experiments in cryogenic environments will be essential to gain deeper insight into the underlying mechanisms of slower cooling.

\subsection{Oscillatory Spectrum}

To investigate the anharmonicity of the pseudo-potential, we obtained the oscillatory spectrum of trapped electrons. As shown in sequence (iii) of Fig.~\ref{Fig3}, we varied the trap MW amplitude to sweep the electrons' COM frequency across the readout mode's resonant spectrum. Since frequency shifts in Eq.~\ref{amplitude_dependent_shifts} predict a decrease in electron frequency due to heating associated with the frequency modulation, we first reduced the trap MW amplitude by $10\,\%$ ($\approx 62\,\mathrm{MHz}$ shift), and then increased it by $20\,\%$ ($\approx 124\,\mathrm{MHz}$ shift), to ensure the entire spectrum was captured within the measurement range.

Figure~\ref{Fig4}(b) presents the observed resonant excitation of the readout mode as a function of the COM frequency sweep. The oscillatory spectrum of trapped electrons exhibited significant distortion, confirming strong anharmonicity in the pseudo-potential. The spectral peak consistently shifted above the initial trap MW amplitude, indicating electron heating, as expected. The Gaussian fitting to the observed spectrum revealed its frequency width to be approximately $109\,\mathrm{MHz}$, substantially larger than the theoretical prediction. A portion of this broadening can be attributed to the high temperature of trapped electrons. Assuming thermal equilibrium at $300\,\mathrm{K}$ of the stationary state in Fig.~\ref{Fig3}(b), the peak height of the oscillatory spectrum suggests that electrons were approximately at $2,950\,\mathrm{K}$ $(0.2\,\mathrm{eV})$ in Fig.~\ref{Fig4}(b). Under these conditions, the estimated frequency broadening at room temperature is about $ 11 \, \mathrm{MHz}$, still significantly exceeding the predicted $2\,\mathrm{MHz}$. This discrepancy could stem from electrode misalignment or stray charge effects.

Based on these measurements, we estimated the number of trapped electrons, neglecting electron-electron interactions. Calibration results indicate that at $300\,\mathrm{K}$, thermal noise from intracavity photons accounts for $87\,\%$ of the noise floor in Fig.~\ref{Fig3}(b). Given this, the signal-to-thermal noise ratio 
 of image current detection is $6.8$, which corresponds to the number of trapped electrons inside a harmonic potential. However, signal degradation due to anharmonicity must be considered. The estimated $11\,\mathrm{MHz}$ broadening reduces the signal to $0.62\,\%$ of its original value, leading to an approximate electron number of $1,260$ in Fig.~\ref{Fig3}(b). We note that this estimation assumes electrons remain stationary at $300\,\mathrm{K}$. However, the actual number could be lower if additional heating mechanisms, such as micromotion heating, are present and the stationary temperature is higher.

\section{Conclusions and Outlook}
We demonstrated the image current detection of electrons in a Paul trap at microwave frequencies, observed their signal decay due to resistive cooling, and obtained their oscillatory spectrum. Our results show that approximately $1,260$ electrons could be trapped for over ten seconds. However, their behavior varies between experimental cycles, likely due to strong anharmonic effects from the pseudo-potential and the large thermal spread of electrons at room temperature. The electron Paul trap at cryogenic temperature is essential to achieve more stable confinement and further systematic studies.

The next challenge toward realizing electron qubits in Paul traps is the image current detection and spin-readout of a single electron at cryogenic temperature \cite{Peng2017}. However, the anharmonicity in the trap potential remains a major obstacle. In Penning traps, this issue is mitigated through tailored static electric field compensation \cite{Fan2023}. In contrast, applying such compensation in Paul traps has been challenging due to the complexity of the nonlinear Mathieu equation and limited to systems with small stability parameters ($ q \ll 0.1$) \cite{home2011}. Recently, we proposed a novel approach that extends static electric field compensation to nonlinear Mathieu equations, enabling anharmonicity suppression of Paul traps even in high stability parameters ($q > 0.5$) \cite{Kento2025}. Implementing this method could improve electron confinement and facilitate single-electron detection and spin-readout, bringing us closer to realizing electron qubits in Paul traps.

\section{Acknowledgments}
The authors thank the Hartmut Lab at University of California Berkely, Ferdinand Schmidt-Kaler at Mainz University, Xing Fan at Harvard University, Ippei Nakamura, Yuta Tsuchimoto, Markus Fleck, Shotaro Shirai, and JaEun Kim for the fruitful discussion. This research was supported by JST PRESTO (Grant No.JPMJPR2258), JST SPRING (Grant No.JPMJSP2108), and ASPIRE (Grant No.JPMJAP2428).

\bibliography{apssamp}% Produces the bibliography via BibTeX.

\end{document}